\documentstyle{mn}
\newcommand{\zi}{ \mbox{\boldmath$\xi$} }
\newcommand{\Bl}{ \left( }
\newcommand{\Br}{ \right)}

\title{Regular and chaotic motion in softened gravitational systems}
\author[Amr A. El-Zant]
       {Amr A. El-Zant \\
        Department of Physics and Astronomy, University of Kentucky, 
Lexington KY 40506, USA
\thanks{Currently at the Center for 
Astrophysics and Space Astronomy, University of Colorado, Boulder}
}
\date{Accepted ....
      Received ....;
      in original form ....}

\pagerange{\pageref{firstpage}--\pageref{lastpage}}
\pubyear{1994}

\input epsfig      

\begin{document}

\maketitle

\label{firstpage}

\begin{abstract}
The stability of the dynamical trajectories of softened spherical 
gravitational systems is examined, both in the case of 
the full $N$-body problem and that of trajectories moving
in the gravitational field of non-interacting background particles.
In the latter case, for $N \ge 10000$,
some trajectories, even if unstable, had exceedingly
long diffusion times, which correlated with the characteristic 
e-folding timescale of the instability. For trajectories
of  $N \approx 100000$ systems this timescale could be arbitrarily 
large --- and thus  appear to correspond to regular orbits. 
For centrally concentrated systems, low angular momentum trajectories 
were found to be systematically more unstable. This phenomenon is 
analogous to the  well known case of trajectories in generic 
centrally concentrated non-spherical smooth systems,
where eccentric trajectories are  found to be chaotic.
The exponentiation times also  correlate with the 
conservation of the angular momenta along the trajectories. 
For $N$ up to a few hundred,
the instability timescales of $N$-body systems and their 
variation with particle number are similar to those of the most
chaotic trajectories in inhomogeneous non-interacting 
systems. For larger $N$ (up to a few thousand) the values of the 
these timescales were found to saturate, increasing significantly
more slowly with $N$. We attribute this to collective effects in the
fully self-gravitating problem, which are apparent in the time-variations
of  the time dependent Liapunov exponents.  
The results presented here go some way towards 
 resolving the long standing apparent paradoxes concerning the 
local instability of trajectories. This now appears  to be a
manifestation of  mechanisms driving evolution in gravitational 
systems and their interactions --- and may thus be a useful diagnostic
of such processes.

\end{abstract}


\section{Introduction}

A persisting and perplexing question regarding the dynamics of $N$-body systems
concerns the interpretation of the local divergence of their trajectories
and the origin of this instability. Does this phenomenon, first pointed out
by Miller (1964), tell us anything non-trivial about the behaviour of 
gravitational systems? Does it, for example,
have  any relation with processes, whether
collective or local, leading to evolution in these systems? 
Is it useful in understanding and characterising such processes?

A series of studies by Kandrup and collaborators 
(e.g., Kandrup et al. 1994) have shown the
local instability of trajectories to be an extremely  robust phenomenon ---
occurring for all initial conditions and perturbations investigated, for a
variety of initial macroscopic configurations, and with an exponential timescale
that is roughly  independent of  these properties and of $N$, 
whether the system is initially in dynamical  equilibrium or not. 
In potentials resulting from collections of
 fixed point particles the exponential divergence
and its constant e-folding time have been checked for $N$ up to a million 
particles (Valluri \& Merritt 1999). 
However, as might be expected, conservation of the
action variables, for systems with separable smoothed out potential, improves
with increasing $N$. Thus, at least in some respects and over 
sufficiently short time-scales, the trajectories of
particles in high-$N$ discrete systems  resemble  
 those in continuous 
ones --- since, qualitatively, the motion is completely determined by those
action variables. This implies that time averages of trajectories 
in steady state discrete systems will
approximate those of the corresponding 
smooth systems for times much larger than the exponentiation time-scale
(which is a fraction of the dynamical time). Nevertheless, the latter do
not exhibit exponential divergence of nearby 
trajectories  in the case of separable potentials.
Moreover, trajectories in fixed smooth potentials are characteristics of the 
time-independent
collisionless Boltzmann equation (CBE). If therefore, as is often assumed,
large $N$-body gravitational systems are adequately described by the steady
state  CBE for timescales long compared to their dynamical time, the 
the phenomenon of the $N$-invariant e-folding time  must be considered
paradoxial.

It can be shown (e.g., Braun \& Hepp 1977; Spohn 1980) that,
for large   large-$N$ gravitational systems,
the dynamics described by the full equations of motion, 
converge towards that resulting from solutions of the 
CBE, provided that the 
potential has bounded first and second derivatives 
(a condition which enters, for example, in the form of  Eq. 2.52 of Spohn). 
This amounts to  ``softening'' the potential to get rid of the singularity at 
the origin. Indeed, in potentials with a short range cutoff, the force 
 due to the mean field 
is always greater than the force between any two particles for large enough
$N$. For example, in a spherical system, 
this simply requires that for a test particle at radius $R$
\begin{equation}
N > \frac{R^{2}}{\epsilon^{2}},
\end{equation}
where $\epsilon$ is the (Plummer) softening length. Or equivalently that
$\epsilon > R/\sqrt{N}$. This in fact justifies the mean field approach embodied 
in the CBE. 

Because, in the case of point mass systems, 
 the binary force between neighbours can be 
larger than the force due to the mean field, formally speaking, the
 dynamics of even a large-$N$ system
is not equivalent to the one described by CBE --- and thus
the exponential divergence  does not in itself signal 
any inconsistency. Perhaps the only puzzling point then is why the timescale
of the exponential divergence does not correlate with particle number while
other dynamical properties of gravitational systems obviously do. 
(As noted above, this is even true  for trajectories of systems of fixed point particles,
where the exponential instability cannot be attributed to any time-dependence:
Valluri \& Merritt 1999).

Goodman, Heggie \& Hut (GHH) conducted an extensive
investigation of the exponential instability and concluded that it is 
 precisely at the  radius where the mean field force and the force due 
to near neighbours 
become comparable that the  latter interactions have the greatest 
contribution to 
the process. Since the interparticle spacing, projected on two 
dimensions, decreases as $1/\sqrt{N}$, a test particle
is likely to suffer one such  encounter 
at every crossing of the system --- independent of $N$. For larger $N$
these encounters last for shorter times and thus the deflection angle 
decreases (roughly as $1/\sqrt{N}$). Nevertheless, the {\em initial} divergence
between two trajectories (in the linear approximation) is independent of 
$N$ (see GHH). Thus the
trajectories of two particles,
 initially separated by a small distance compared to the 
impact parameter, will diverge exponentially with a timescale independent 
of $N$. 
Consider however two trajectories separated in such a way that one
is affected by an encounter with a field star at impact parameter
$\sim R/\sqrt{N}$ and the other not significantly affected --- being at a 
distance  $\gg R/\sqrt{N}$ from the perturbing object.
The affected test particle will
be deflected by an angle $\theta \sim 1/\sqrt{N}$. Thus, for large 
(compared to $R/\sqrt{N}$)
separations and large $N$, the divergence will be small.

In conventional dynamical  systems
where chaos is present for most initial 
conditions, it is usually either 
 produced by the bulk properties of the potential 
or by local large angle scatterings.
In both cases large deviations of trajectories, resulting from the 
non-integrable component of the potential,
are expected
in times comparable to the exponentiation times.
In unsoftened gravitational systems on the other hand, the contribution
to the potential responsible for exponential divergence on very 
short timescales independent of $N$ 
comes from encounters whose range, duration, and deflection angles
progressively become smaller at larger  $N$.  
Their effect on the actual trajectories becomes small,   
and the divergence does not lead to  any macroscopic evolution on 
the divergence timescale  --- since it is only at small separation that the 
divergence is independent of $N$.
Any evolutionary effects due to the exponential instability, therefore,
will have to come from larger range interactions, whether individual or 
collective.

In order to answer the questions posed in the opening paragraph
of this paper one will therefore need to eliminate effects due to encounters 
in the interaction range which contributes to the $N$-invariant e-folding time,
but do not, for sufficiently large-$N$, significantly change particle 
trajectories on such timescales.
When this is done, the mathematical conditions for large $N$-body systems 
to be described by the CBE, for times much longer than the dynamical time, 
are then also satisfied. 
An important consequence therefore is, 
if these large-$N$ systems  are adequately described by 
steady state solutions of the CBE, the divergence timescale 
in systems where the smoothed out  background potential is separable 
will have to increase with $N$. Moreover, since in such a description 
there are no particle-particle correlations and trajectories are 
independent, divergence  timescales in  the full $N$-body 
problem and their variation with $N$ should be similar 
to those of trajectories in compatible systems
of non-interacting background particles. It is the object of this 
paper to examine the validity of such assertions.

 By introducing a short range cutoff, GHH studied the behaviour 
of the divergence timescale due to random long
(compared to $R/\sqrt{N}$) range interactions. They found it to increase as 
$N^{1/3}$.  Calculations of the Liapunov exponents 
from full $N$-body simulations of softened particles  over short 
timescales confirmed this for low $N$ (up to a few hundred). 
One of the goals  of the present paper is the 
the investigation of the variation of the exponentiation times of 
softened $N$-body systems  to higher $N$ and 
longer times (Section~\ref{nbody}). We confine ourselves to spherical 
systems which, being separable in the continuum limit, should have, in that limit,
 steady states
completely characterised by regular orbits with no exponential divergence. 
In addition, in an attempt to isolate effects that depend on the
full self-consistent $N$-body 
interaction from those associated with  discreteness noise resulting 
from rapid spatial and time variations of the potential, 
we also
examine the divergence of single trajectories moving in systems of 
non-interacting softened particles. We do this for three different
configurations: a collection of non-interacting particles moving in a 
spherical box (Section~\ref{nonmov}),
 a statistically homogeneous distribution of 
fixed particles,  and a distribution of  particles with 
density  decreasing with radius as $1/r^{2}$ (Section~\ref{nonfix}). 
This latter
system is known to contain chaotic orbits when non-spherical 
perturbations are introduced, even in the continuum limit. 
We here examine
whether this feature leads to different stability properties for its 
trajectories in comparison with the homogeneous case (where, in the 
aforementioned limit, the potential is harmonic and all trajectories are
regular even in the triaxial case), in the hope of exploring the effects
of the smoothed out mass distribution on the stability 
of trajectories in discrete systems.  
We summarise and attempt to interpret the results in 
Section~\ref{conremks}.

\section{$N$-body systems}
\label{nbody}
\subsection{Numerical evaluation of Liapunov exponents}

Liapunov exponents measure the linear stability along the trajectories of 
dynamical systems. For example, let
\begin{equation}
{\bf \ddot{x}_{i}=f_{i}} 
\end{equation} 
be the Newtonian equations of motions of an $N$-body gravitational 
system (with $i=1,N$), and 
\begin{equation}
{\bf \delta \ddot{x}_{i}=\delta f_{i}}                                                                         
\end{equation}
the corresponding variational (i.e., linearised) equations. 
The latter measure the deviation of 
nearby states to the one determined by the first set of equations. Next
define
\begin{equation}
{\bf X=(x_{1},...,x_{N},\dot{x}_{1},...,\dot{x}_{N})}
\end{equation}
as the $6N$-dimensional ``vector field'' of the system,   and
\begin{equation}
{\bf \mbox{\boldmath$\xi$}=(\delta x_{1},...,\delta x_{N},
\delta \dot{x}_{1},...,\delta \dot{x}_{N})}
\end{equation}
as the corresponding vector field of variations. Then the 
above equations can be written as 
\begin{equation}
{\bf \dot{X}=F}       \label{eq:linf1}
\end{equation}
and
\begin{equation}
{\bf \dot{\mbox{\boldmath$\xi$}}=\delta F} \label{eq:linf2}.
\end{equation}
Now if we choose a particular trajectory  
${\bf \bar{X}=\bar{X}}(t,t_{o},{\bf \bar{X}_{o}})$ against which we would
like to measure the deviation, with ${\bf X_{o}}$ being the point
where we start measuring that deviation (i.e., the initial conditions), 
then~(\ref{eq:linf2}) can be
rewritten as
\begin{equation}
\dot{\mbox{\boldmath$\xi$}}={\bf D_{x}F(\bar{X}}(t,t_{o},{\bf X_{o}}))
    \mbox{\boldmath$\xi$},  
\label{eq:linf3}
\end{equation}
where ${\bf D_{x}F}$ is the Jacobian $6N \times 6N$  matrix \hspace{0.05in}
${\bf \partial F_{i}/\partial x_{j} }$ \hspace{0.05in} and  $i,j=1,6N$. 
Now let
\begin{equation}
{\bf X_{s}=X_{s}(\bar{X}}(t,t_{o},{\bf X_{o}}))
\end{equation}
 be the fundamental
solution of this matrix with the initial condition being the identity
matrix, the solution of~(\ref{eq:linf3}) is then given by (Wiggins 1991)
\begin{equation}
      \mbox{\boldmath$\xi$}={\bf X_{s}}(t)\mbox{\boldmath$\xi_{0}$},
\end{equation}
which describes the evolution under the linearised dynamics with initial conditions
$\mbox{\boldmath$\xi_{0}$}$ in the space of linear variations.

A Liapunov exponent is the infinite limit of the 
``time dependent Liapunov exponent'' (Wiggins 1991)
 at ${\bf X_{0} }$ in the direction 
${\mbox{\boldmath$\xi_{0}$}}$ at time $t$ which
is given by
\begin{equation} 
\lambda(\zi,t)=\frac{\parallel
\zi(t)\parallel}{\parallel \zi_{0}
\parallel}= \frac{1}{t} \log \Bl \frac{\parallel{\bf X}_{s} (t)\zi
\parallel}{\parallel \zi_{0} \parallel}\Br \label{eq:linf4}.
\label{epn}
\end{equation}
The Liapunov exponents are then defined as
\begin{equation}
\sigma(\zi_{0}, {\bf X}_{0})=
\lim_{t \rightarrow \infty} \lambda(\zi_{0},{\bf X}_{0},t) 
\label{eq:linf5}.
\end{equation}
Numerically of course only $\lambda$ can be calculated. 
We will refer to the inverse of this time dependent Liapunov 
exponent as the ``exponentiation time'', the ``exponential 
timescale'' or the e-folding time.

For a Hamiltonian
system with $f$ degrees of freedom,
there are $2f$ linearly independent directions in  phase space
for the vector $\mbox{\boldmath$\xi_{0}$}$ to point at, hence
there are $2f$ Liapunov exponents. A positive Liapunov exponent indicates
unstable behaviour characteristic of chaotic motion. Thus, determining the maximal
exponent is sufficient for detecting the  presence of such behaviour.
The evaluation of the  maximal exponent is straightforward
enough. This is because  exponential instability, if it is present,
will cause almost all initial 
linear tangent space vectors to realign themselves along the subspace
of maximal expansion. A numerical determination of  a Liapunov 
exponent from almost {\em any} initial chosen direction for the linear variations 
will thus tend to give an evaluation of the maximal exponent (Wolf et al. 1985).  
The only complication that arises is that, 
when the exponentially increasing
solutions of the linearised equations become too large,
the calculation is slowed down (eventually leading to numerical 
overflow).  
This is easily remedied however by application
of the ``standard algorithm'' of Benettin et al. (1976). 
This algorithm is based
on the local averaging of the deviation between neighbouring
states, which is done by dividing the time we run the system into $n$
subintervals. An initial linearised deviation $\mbox{\boldmath
$\xi_{0}$}$ will therefore be transformed into 

\begin{math}
\mbox{\boldmath
$X_{s}^{1}\xi_{o},X_{s}^{2}X_{s}^{1}\xi_{o},...,X_{s}^{n}...X_{s}^{2}
X_{s}^{1}\xi_{o}$} 
\end{math}
 at times
$t_{1},t_{2},...,t_{n}$. The right hand side of~(\ref{eq:linf4})
can then be rewritten as:
\begin{equation}
\mbox{\boldmath$\parallel X^{n}_{s}...X^{3}_{s}X^{2}_{s}
X^{1}_{s}\xi_{0}\parallel/\parallel\xi_{0}\parallel$}. 
\end{equation}
We now successively define  
\begin{equation}
\mbox{\boldmath$ \xi_{i}=X^{i}_{s}\xi_{i-1}=\parallel \xi_{i-1}
\parallel X^{i}_{s}$}\hat{\mbox{\boldmath$\xi$}}_{i-1}
\end{equation}
with  
\begin{equation}
\hat{\mbox{\boldmath$\xi$}}_{i-1}=\mbox{\boldmath$\xi_{i-1}/\parallel \xi_{i-1}\parallel$}
\end{equation}
this means that 
\begin{equation}
\mbox{\boldmath$ \parallel X_{n}...X_{2}X_{1}
\xi_{0}\parallel=\prod_{i=1}^{n}\parallel \xi_{i}\parallel $}
\end{equation}
and therefore 
\begin{equation}
\lambda_{standard}=\lim_{n \rightarrow \infty}\sum^{t/\Delta t}_{1}
\frac{\log \parallel \mbox{\boldmath$\xi_{i}$}\parallel}{t}. \label{eq:linf6}
\end{equation}
In practice, this procedure consists of renormalising the linearised vector 
to unity
at intervals $\Delta t$, adding the logarithm of its norm to the pre-existing 
sum and
restarting the integration with this renormalized unit vector serving 
as initial condition
for the variational (linearised) equations. This avoids numerical blowup.

\subsection{Parameters and initial conditions}

\begin{figure*}
\epsfig{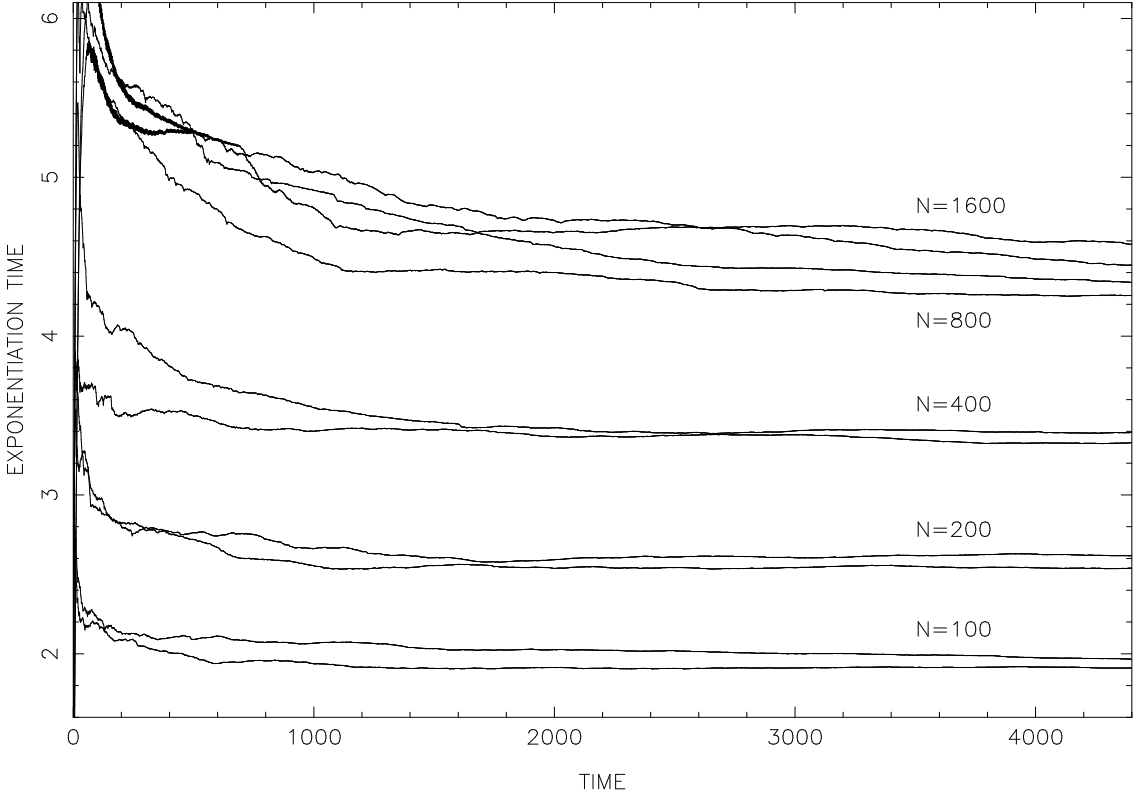}
\caption{Time variation of inverse of the time dependent Liapunov exponents 
(calculated from Eq.~\ref{epn}) for different particle numbers and random 
realisations of self gravitating $N$-body systems.}
\label{eponetials3}
\end{figure*}

\begin{figure*}
\epsfig{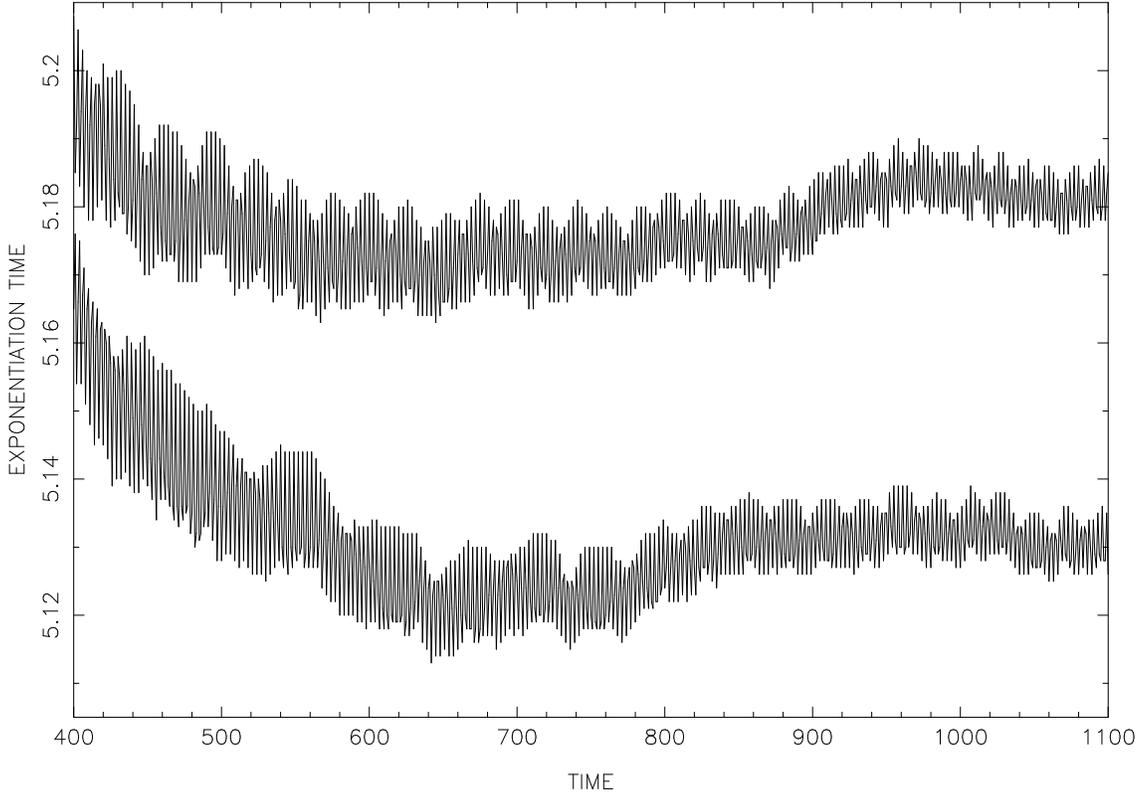}
\caption{Time variation of the inverse of the time dependent Liapunov exponents 
 for two different realisations of $N=3200$ 
self gravitating systems. After the initial evolution. leading to a more or less 
quasi-steady state, weakly damped modes
can remain for at least a few hundred dynamical times.
}
\label{wavesplot}
\end{figure*}

In general, the values of the 
 Liapunov exponents will depend on the initial conditions.
If the phase space is ``topologically transitive'' however --- that is if 
trajectories can visit any region of the phase space --- for almost all 
initial conditions in the connected region all the Liapunov
 exponents will be equal. This however is not true of the time dependent Liapunov
exponents which can be different at any given time.  
Convergence indicates that an invariant phase space distribution has been 
reached. The convergence time is therefore a measure of the
transport properties of a given system's phase space. 
The time dependent Liapunov exponents of systems with more complex phase
space take longer to converge. Liapunov exponents of regular orbits 
converge to zero at a rate $\sim \frac{\ln t}{t}$.

Strictly speaking, Liapunov exponents 
of open gravitational systems do not converge at all, because 
these systems are thermodynamically unstable and continuously evolve 
towards more and
more concentrated states 
(e.g., Antonov 1962; Lynden Bell \& Wood 1968; Padmanabhan 1990). 
Moreover, the evolution rate will depend on $N$, making the 
comparison of  systems with different $N$ ambiguous.
This is not the 
case however for enclosed $N$-body systems of single mass particles started 
from appropriate initial conditions and whose energy, mass and radius
satisfy (El-Zant 1998)
\begin{equation}        
ER/GM^{2} > - 0.335 .
\label{colcon}
\end{equation}

We define our system parameters such that
$G=M=R=1$, randomly distributing our particles inside $R=1$.
The quantity on the left hand side of 
condition~(\ref{colcon}) is then solely determined  by the initial velocities.
These are  taken to be  random in direction and their absolute values
vary with radius as $v=v_0 \exp (r)$, where $v_0$ is determined by the 
initial kinetic energy of the system. Since we have already determined the 
spatial distribution, this amounts to fixing the virial ratio.  Our systems 
must have virial ratios greater than $0.5$ if they are to satisfy 
relation~(\ref{colcon}) 
and be thermodynamically stable. At the same time the virial ratio 
should not be too large, so as not to modify the dynamics significantly.
We choose a virial ratio of $0.69$. Systems starting with these initial 
conditions were shown (El-Zant 1998) to quickly evolve towards 
quasisteady isothermal 
distributions, with the total change in the virial ratio 
throughout the evolution being of the order of  a few percent.
Systems are enclosed using the method also described in  
 El-Zant(1998): particles
venturing beyond $R=1$ are subject to a force of the  form $m K (1-r)^{3}$, 
with $K=1300$ and $m=1/N$ is the mass of a particle.

We numerically integrate the Newtonian equations of motion along 
with their variational counterparts starting with random initial 
conditions for the latters (each component of the
vectors in the tangent space of variations takes a value randomly
chosen between zero and one and the vector is normalised so that 
its norm is one). In the simulations presented here
the boundary force is not included in the variational equations, 
however its inclusion was not found to significantly affect the results. 
The integration is advanced using a highly accurate
variable order variable stepsize Adams method, as implemented in the NAG
routine D02CJF with tolerance $10^{-5}$. 
For systems  consisting of a $100$ 
particles,  the dependence of the macroscopic evolution on the tolerance
was checked and found not to vary significantly for tolerances of 
$10^{-3}$ to $10^{-10}$. For the tolerance used here, 
the energy was conserved to 
better than one part in 10000 for a hundred units and better than one 
percent for 20000. The energy conservation is much better for larger 
particle  numbers.
The renormalisation time for the variational
equations  was taken to be one unit. For the above parameters
this is of the order of a  dynamical time. More precisely, if one
defines the crossing time of a system in virial equilibrium by
\begin{equation}
T_{cross}= M^{5/2}/ (2 E)^{3/2},
\end{equation}
it is found to be about 4 time units for systems with the above parameters.
In practice because our systems have a virial ratio of $0.69$ instead of
$0.5$, the crossing time is $0.67$ times shorter.
In all simulations the (Plummer) softening of length is fixed at $0.1$
($ \ge R/\sqrt{N}$ for all $N$ investigated).

\subsection{Results}

For the full $N$-body problem, a total of 13  runs were investigated. 
Two different realisations  for each  $N=100, 200, 400, 800$ system, 
three for $N=1600$ and four with systems of  $N=3200$ particles.
The $N=100$ runs were integrated up to $20000$ time units and so was one
of the $N=200$ runs, the other one was stopped at $N=6600$ when it was clear 
the results were converging. The $N=400$ runs were integrated up to $7000$
time units and the $N=800$ for $5000$ and $10000$ units. For $N=1600$,
two runs were pursued until $4400$ units while one was stopped at 
$1500$ units. Finally, two of the $N=3200$ runs were integrated up to $700$
units and the other two for an additional $400$ units, for a total of
1100 time units. 

\begin{figure}
\epsfig{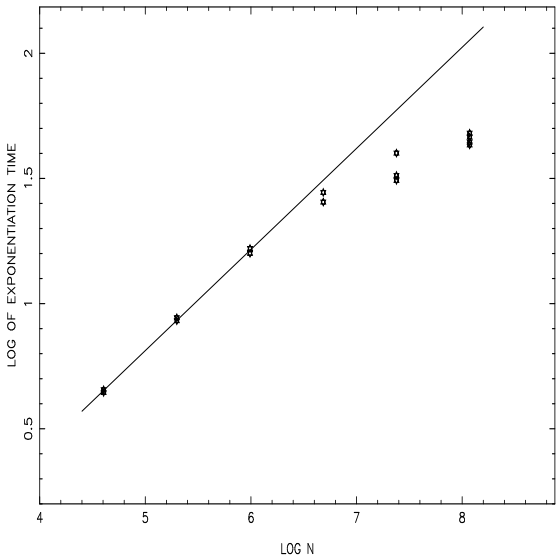}
\caption{Variation of the exponential timescales with particle number for 
self gravitating $N$-body systems. The plotted line, which fits well the results
for runs with $N=100, 200$ and $N=400$ has slope of $0.4$}
\label{fullplot}
\end{figure}

\begin{figure*}
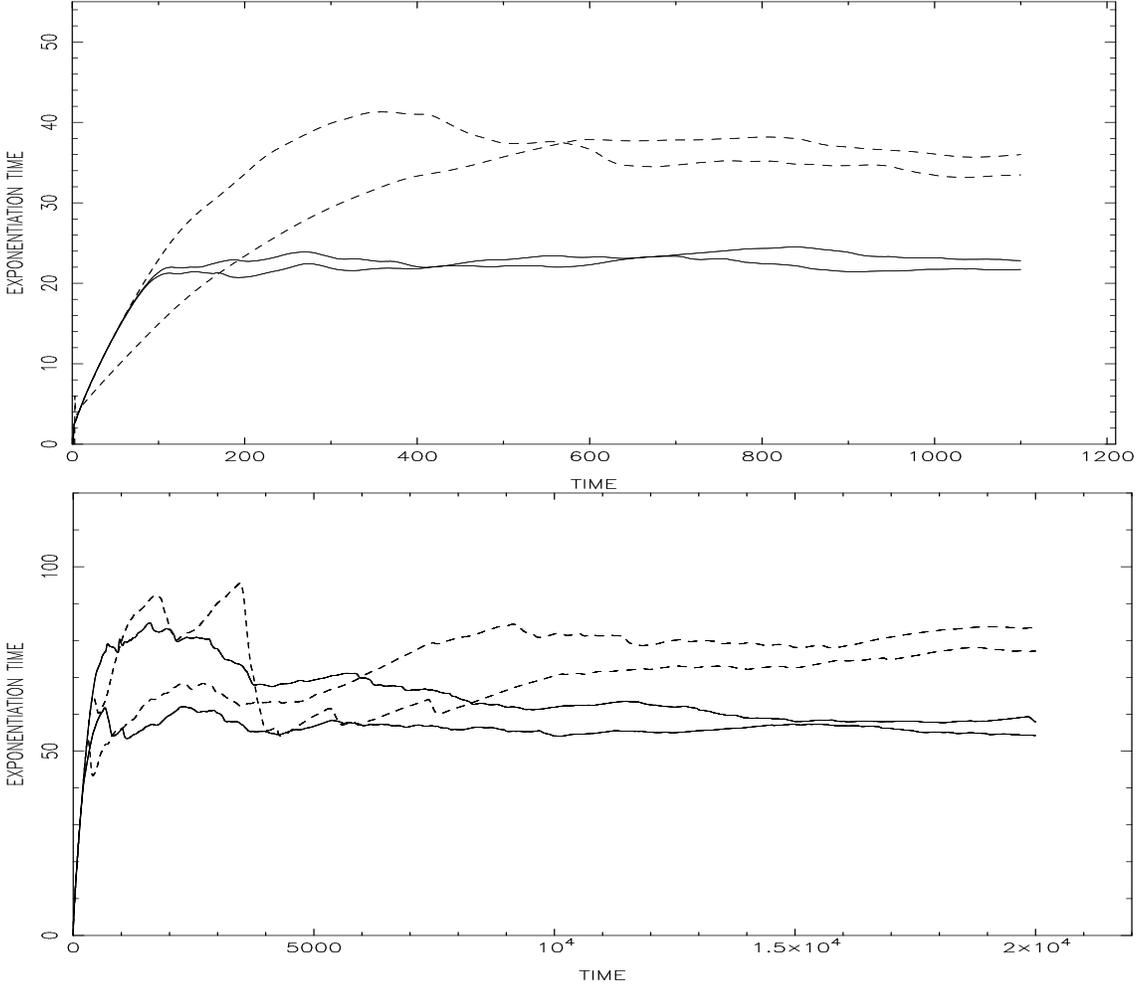

\epsfig{file=shortplots.ps,width=15.cm,height=6.5cm,angle=-90}
\epsfig{file=longplots.ps,width=15.cm,height=6.5cm,angle=-90}
\caption{Time variation of the inverse of the time dependent Liapunov exponents 
(calculated from Eq.~\ref{epn}) for systems where one (test) particle interacts with 
all the other  (background) particles, which are non-interacting. The upper panel
shows results for random initial realisations of systems consisting of   $1000$ and $2000$ 
particles, while the lower panel shows the corresponding results when $N=4000, 8000$.
Invariably, the final exponentiation times are longer for systems with larger $N$. 
Convergence however is much slower for larger systems (note the different time scales
in the upper and lower plots).
}   
\label{shortlongs}
\end{figure*}

As can be seen from  Fig.~\ref{eponetials3}, for simulations that were 
run up to $4400$  time units, the exponential divergence timescales show 
good convergence, although there seems to be a small systematic decrease in 
some of the higher-$N$ runs. This appears to be a result of slight increase in central 
concentration accompanying slow evolution towards the invariant equilibrium state
(reached much quicker in the lower-$N$ runs).
Also, in these runs, it is possible to detect 
small regular oscillations in the exponential timescales for early times 
(smaller than $800$ units). These are eventually damped out (although
variations on longer timescales persist for much longer).
 For $N= 3200$  however, these oscillations are much larger and are not
completely damped over the timescales investigated (Fig.~\ref{wavesplot}).
These oscillations reflect weakly damped  global density oscillations. 
 By inspection, at least
three timescales can be detected. They are of the order of one, ten and few 
tens of time units.

Already from  from Fig.~\ref{eponetials3} and Fig.~\ref{wavesplot}, it is
apparent that the
exponentiation  timescale increases with $N$, but the rate of this increase
 decreases for larger $N$.
This is evident in  Fig.~\ref{fullplot}, where we have plotted the final
exponentiation times for all the runs as a function of $N$. The straight 
line in that figure fits the first three values of $N$. It has a slope 
of $0.4$, which is roughly compatible with a slope of a third, predicted
by GHH on the basis of their analysis of the divergence in a infinite
homogeneous medium of softened particles,  and confirmed by their simulations at 
lower $N$. 
One therefore naturally suspects that the discrepancy results from
self gravitating modes leading to fluctuations in the mean
field, as evidenced by the oscillation of the values of the exponential 
times  for higher values of $N$  noted above. These global oscillations 
appear to be quickly washed out by discreteness noise 
in the low $N$ ($ < 800$) limit.

For purposes of comparison, in order to investigate further 
the possibility that  effects due to self-gravity are indeed responsible 
for the saturation of the exponentiation times, we consider in the 
 next sections several  systems where self gravity is not included.

\section{Non-interacting moving particles}
\label{nonmov}

\begin{figure}
\epsfig{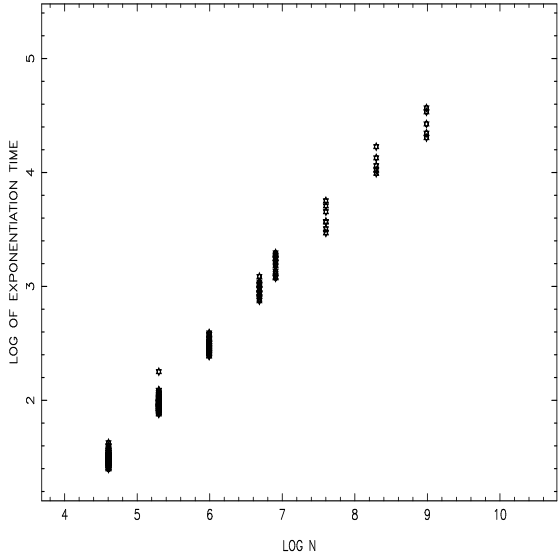}
\caption{Variation of the exponential timescales with particle number for 
systems similar to those described in Fig.~\ref{shortlongs} (see also text).
Least square fits result in power laws with index $0.64$.
}
\label{oneplotallcomp}
\end{figure}

In this approximation one particle interacts with all other particles, 
which do not interact with each other (but the motion of which is affected 
by the interaction with that particle). 
The system is enclosed as in the 
previous section and the total kinetic energy is scaled so as to equal 
that of the self-gravitating systems described there. The interacting 
particle is picked randomly from the spatially homogeneous initial
particle distribution, and the full set of $12 N$ first order differential
equations, describing the motion of all $N$ particles and 
its variations, is integrated. The Liapunov exponents are then obtained in the 
same manner as in the previous section (with the force law now only containing 
the interactions of one particle with all others).
All system parameters and units (including the softening) are the same as 
before.

We have run simulations with 
$N=100, 200, 400, 800, 1000, 2000, 4000$ and $8000$.  
For $N= 100, 200$ and $400$ respectively,
we integrated  $100$, $88$ and $38$  initially statistically identical 
systems for $1100$ time units. For $N=800$ and $N=1000$ we integrated 
$16$ such systems for the same time interval. For $N=2000$ there were 
five runs that went for times up to $1100$, one up to $6600$ and  
four up to $25000$ time units. Five  $N=4000$ runs were completed, the first
three were integrated for $8400, 8800$ and $10000$ time units. These were supplemented
by two $20000$ time unit runs. Finally, for $N=8000$, there were also five
runs, three for  $4000$ units and two for $20000$.

Since the field particles are moving, the test particle's energy 
is not conserved. In principle therefore, the whole allowed phase space 
is shared by any two initial conditions, provided that the motion is 
ergodic. The Liapunov exponents (and the associated exponentiation times)
 should therefore converge to the same 
asymptotic values  for all initial conditions. 
Because  the background particles in this system are 
non-interacting, no global modes can be present and the system must be in a statistically
steady state (characterised by a homogeneous density distribution).
Evolution is then driven by discreteness noise alone, the power of which decreases 
with $N$.  
Diffusion timescales, therefore, should be longer for larger $N$, and the 
convergence timescales should then also scale accordingly ---
since longer time is needed to reach an invariant phase space distribution. 
This was indeed
found to be the case, as is illustrated in Fig.~\ref{shortlongs}, where 
we show the  
inverse of the time dependent Liapunov exponents
  for pairs of runs with different 
particle numbers ($1000, 2000, 4000$ and $8000$). Note that both  the convergence
and exponentiation timescales are much shorter for smaller $N$. 
This is what is to be expected if  the divergence 
timescale is related to ``mixing'' and the loss of memory of initial conditions.
  Nevertheless, for the runs described in this section, convergence
was always reached --- which implies that all trajectories  are 
chaotic and, for a given $N$, share the same region of phase space. As we 
will see in the next section, this does not appear to be always the case
with trajectories of non-interacting softened systems.

  The variation of the final exponential timescales as a function of 
particle number, for all the runs, is shown in Fig.~\ref{oneplotallcomp}. 
As can be seen,  not only is there no clear sign of  saturation in the increase 
of the exponentiation
timescales for the larger $N$ runs, but rate
of  increase in the exponentiation times
is obviously very different from the self gravitating case. The observed 
$N$-variation is very well fit by a power law. Best least square fits
gave a slope of $0.64$ (plus or minus $0.008$). This is rather different 
from the slope of $\sim 1$ expected if there was direct correspondence with 
the $N$ variation of the relaxation time inferred from two body relaxation 
theory --- even though the situation here  is clearly similar to 
the idealised systems that are the subject of 
this theory (see, e.g., Chandrasekhar 1942). It is however much larger than 
the slope of the corresponding relation for the exponentiation timescales in the 
full $N$-body problem --- even for small $N$.

\section{Non-interacting fixed particles}
\label{nonfix}

\begin{figure*}
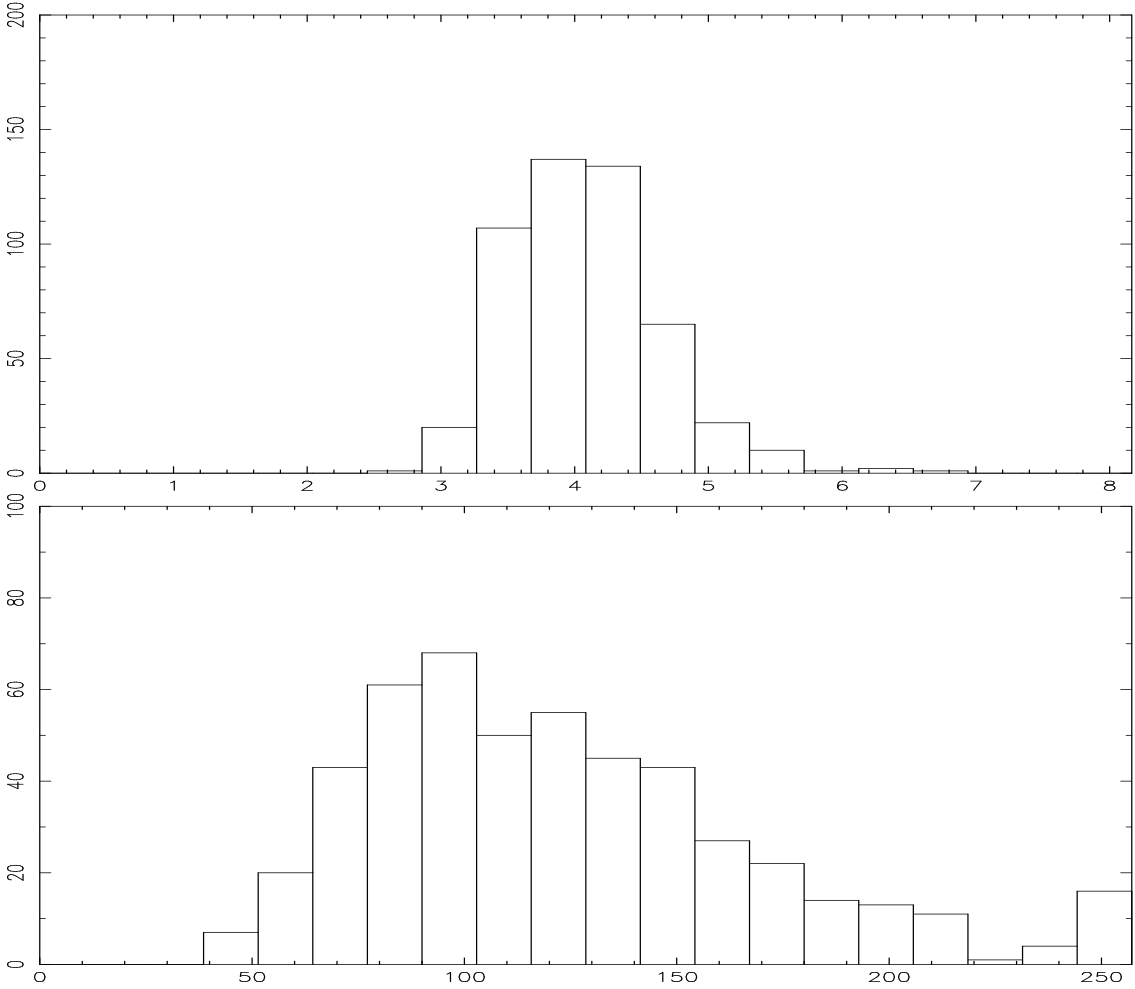

\epsfig{file=histo100.ps,width=15.cm,height=6.5cm,angle=-90}
\epsfig{file=histo10000.ps,width=15.cm,height=6.5cm,angle=-90}
\caption{Distribution of the values of the inverse of the time dependent
Liapunov exponents  
after $420$ time units 
for $500$ trajectories moving in the potential  of 
(constant density) fixed particle
systems of $N=100$ (upper panel) and $N=10000$. In the latter case, the bin centered 
around $250$ also contains all trajectories with larger exponentiation times.
}
\label{histos}
\end{figure*}

\begin{figure}
\epsfig{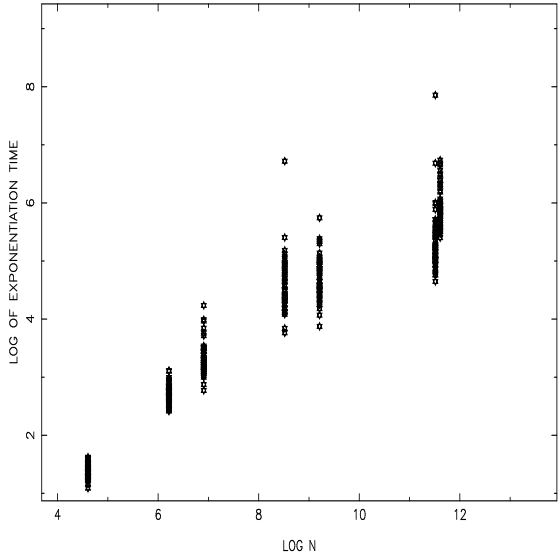}
\caption{Variation of the inverse Liapunov exponents of trajectories of 
homogeneous systems of fixed point particles. All trajectories were integrated  for
$420$ time units, except in the case $N=100000$ where some trajectories were also
integrated for $4200$ units (the exponents for these  are shown slightly to the right
of the $420$ unit ones).
}
\label{fixplot50}
\end{figure}

Integrating the full equations of motion, even for enclosed systems of 
non-interacting particles, for long intervals, is time consuming. To 
investigate the behaviour of the Liapunov exponents for larger $N$ we 
therefore  revert to an even simpler situation, whereas we integrate
individual trajectories  moving in the potential of systems of
{\em fixed} particles for short times but for a large number of initial 
conditions. For each initial condition, only six first order equations of 
motion along with their associated variational equations are integrated.
The Liapunov exponents are then obtained by the same method as in the
previous sections. 

Two types of distributions are discussed: homogeneous 
and $\rho \sim 1/r^{2}$. All system  parameters and units are the same 
as in the previous sections. The velocities
are chosen such that the total energy is smaller than the
energy of the zero velocity surface $V_{ZVS}$ for a corresponding 
smooth distribution at $R=1$
(which, for large $N$,  roughly coincides  with the  zero 
velocity surface corresponding to the actual particle distribution).  
That is, each velocity component $i$ is determined  from 
$ v_{i}= \sqrt{-2 (V+V_{ZVS})} \times (1 - 2 X)$, where $X$ is a random
number uniformly distributed between 0 and 1 and $V$ is the particle's
initial potential energy as determined by its initial spatial 
coordinates. The (Plummer) softening length is again fixed at 0.1.

\subsection{Homogeneously distributed systems}

Fig.~\ref{histos}
 shows histograms displaying the distribution of the exponentiation timescales 
of samples of $500$ trajectories integrated in homogeneous systems 
of $100$ and $10000$ softened particles,  with the initial 
conditions described above
and for $420$  time units. As can be seen, the spread in exponentiation
times {\em increases} significantly for larger $N$. 
No obvious correlation was found between the values of the Liapunov exponents for a 
given $N$ and the initial conditions of the trajectories. In fact, starting 
from the same initial conditions and randomly varying the positions of
the background particles gave similar results. This is indicative 
of a very complex phase space structure that may well
be worth studying in more detail in the future.
Perhaps more important, it may also indicate that a 
 transition from a highly 
chaotic regime with a single  Liapunov exponent to a mixed phase space
is taking place.

In the case of systems without discreteness noise, that is when the potential does not
contain rapid spatial variations as is the case here, mixed phase spaces
have  sets of initial conditions from which orbits have zero Liapunov
exponents (in the infinite time limit). This appears to be the case  
for some of the trajectories of the $N=100000$ systems. For these,  
monotonous decrease of the Liapunov
exponents (roughly with the expected rate of  $\log t /t$) was detected
for times up to $4200$ units. Thus it appears that {\em some trajectories actually
become regular} for $N=100000$, despite the discreteness of the system.
Nevertheless, most exponents tended to converge to some non-zero value when the integration
was continued for longer times. It is possible that Arnold diffusion 
(see, e.g., Lichteberg \& Lieberman 1983), highly ineffective
in potentials  without high frequency variations, is at work here. Since 
the associated diffusion
timescales could be very long, convergence is expected to be slow. 
For higher $N$ convergence 
is slower --- hence the increase in dispersion.

Fig~\ref{fixplot50} displays the the changes in the exponentiation times as 
a function of $N$  
for  subsets of 50 randomly chosen trajectories 
of systems with  $N = 100, 500, 1000, 5000, 10000, 100000$. As expected, one sees an 
increase of the average value of the exponentiation times accompanied by 
an increase in their scatter. Even though it appears there is  some 
saturation in the average values  as $N$ increases, this may be due to
the relatively short integration times (which, as is by now clear, 
can prevent convergence of the Liapunov exponents to their final values). 
For example, in the case of $N=100000$,
we have integrated $50$ randomly chosen initial conditions for ten times
longer than the other trajectories (integrated for the 
standard time interval of $420$ time units). The exponentiation timescales
for these longer time integrations are shown slightly to the right. It is 
clear that they are, on average, larger.
  Best fits
for the values of the exponentiation times as a function of $N$ are 
 compatible with those found in the previous section for the 
non-interacting moving particles.

\subsection{Systems with density proportional to $1/r^{2}$}

As was mentioned in the previous subsection,
in the case of the homogeneous systems studied there,  no correlation 
was found between the values of the exponentiation times and the initial 
conditions.
As  we will now see, this is not the 
case for systems with centrally concentrated particle distributions.

\begin{figure*}
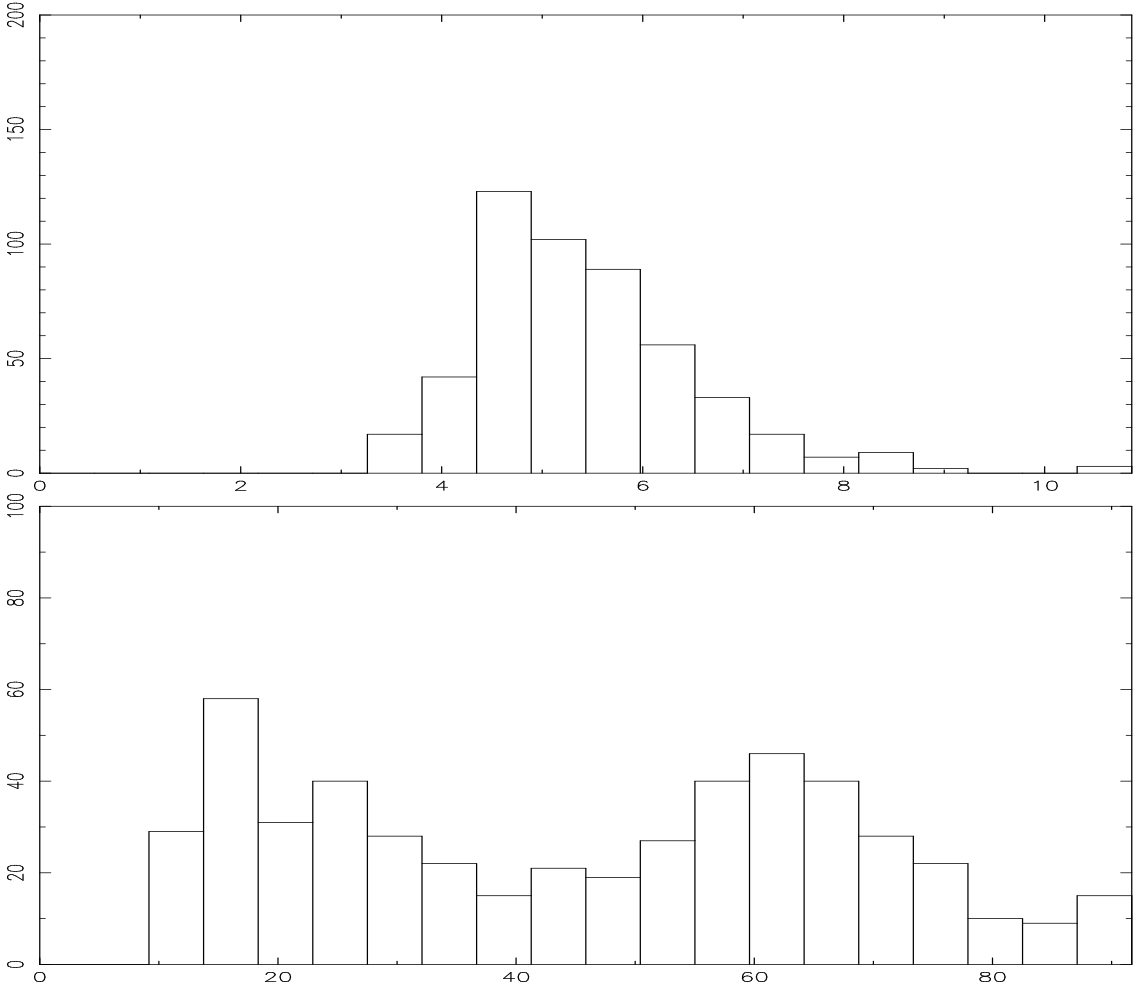

\epsfig{file=histo100r2.ps,width=15.cm,height=6.5cm,angle=-90}
\epsfig{file=histo10000r2.ps,width=15.cm,height=6.5cm,angle=-90}
\caption{Distribution of the inverse of the Liapunov exponents 
after $420$ time units 
for $500$ trajectories moving in the potential  of 
 fixed particle
systems with $\rho \propto 1/r^{2}$
and with $N=100$ (upper panel) and $N=10000$
}
\label{histosr2}
\end{figure*}

\begin{figure*}
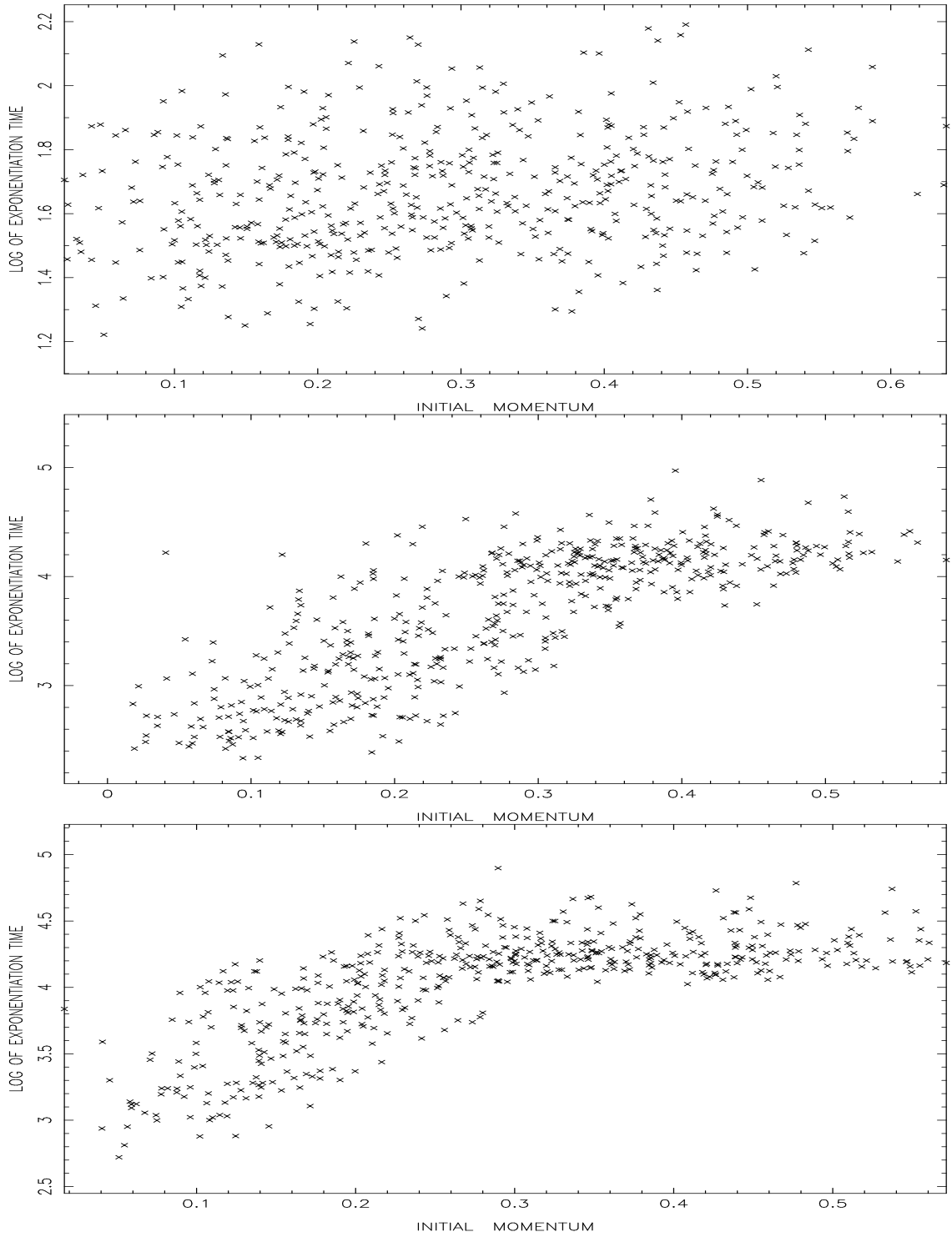

\epsfig{file=momplot100r2.ps,width=15.cm,height=6.5cm,angle=-90}
\epsfig{file=momplot10000r2.ps,width=15.cm,height=6.5cm,angle=-90}
\epsfig{file=momplot50000r2.ps,width=15.cm,height=6.5cm,angle=-90}
\caption{Exponentiation times as as function of initial angular momenta
for  trajectories integrated for $420$ time units in 
 systems of $100$ (upper panel) $10000$ (middle panel)
and $50000$ particles distributed  so that $\rho \propto 1/r{2}$
} 
\label{momplot}
\end{figure*}

Fig.~\ref{histosr2} shows histograms for a sample of $500$  trajectories 
integrated in particle distributions with $N=100$ and $N=10000$ 
and average densities varying as $\sim 1/r^{2}$.
Evidently, the major difference from Fig.~\ref{histos} is that, along with
the increase in the  average value and the scatter of the exponentiation
times, their distribution becomes bimodal. 
These two groups
turn out to be  correlated with two distinct sets of initial conditions
--- 
which in turn can be related to stability characteristics of trajectories 
in smooth non-spherical potentials. 
Recall that in such potentials, in the presence of a central mass 
concentration or a central density cusp, elongated orbits passing
near the centre tend to be chaotic. Most box orbits and elongated
loops with low initial angular momentum (which may also include low angular 
momentum orbits in axisymmetric potentials: e.g., 
Caranicolas \& Innanen 1991)
will fall in this category
(e.g., Gerhard \& Binney 1985; Schwarschild 1993; Merritt \& Fridman 1996;
El-Zant \& Hassler 1998).

Fig.~\ref{momplot} shows the correlation between the initial angular momenta of the
integrated trajectories and the corresponding exponentiation times. As is 
clear, for $N=100$, little correlation exists: the dynamics is dominated by 
the discreteness noise and trajectories can diffuse freely from one distribution
to another. However, for $N=10000$ and $N=50000$, a clear 
correlation emerges between the two quantities, with trajectories starting
with higher values of the initial angular momentum being more regular. 
Thus, in these cases, the discreteness noise acts as a perturbation ---
having  similar effects to those of  non-spherical perturbations 
in smooth systems. In both situations, trajectories starting at higher angular
momenta are relatively immune to the destabilising effect of the 
non-integrable perturbation, as compared to the ones starting from low 
angular momenta (for an explanation of this phenomenon see 
El-Zant \& Hassler 1998).

\begin{figure*}
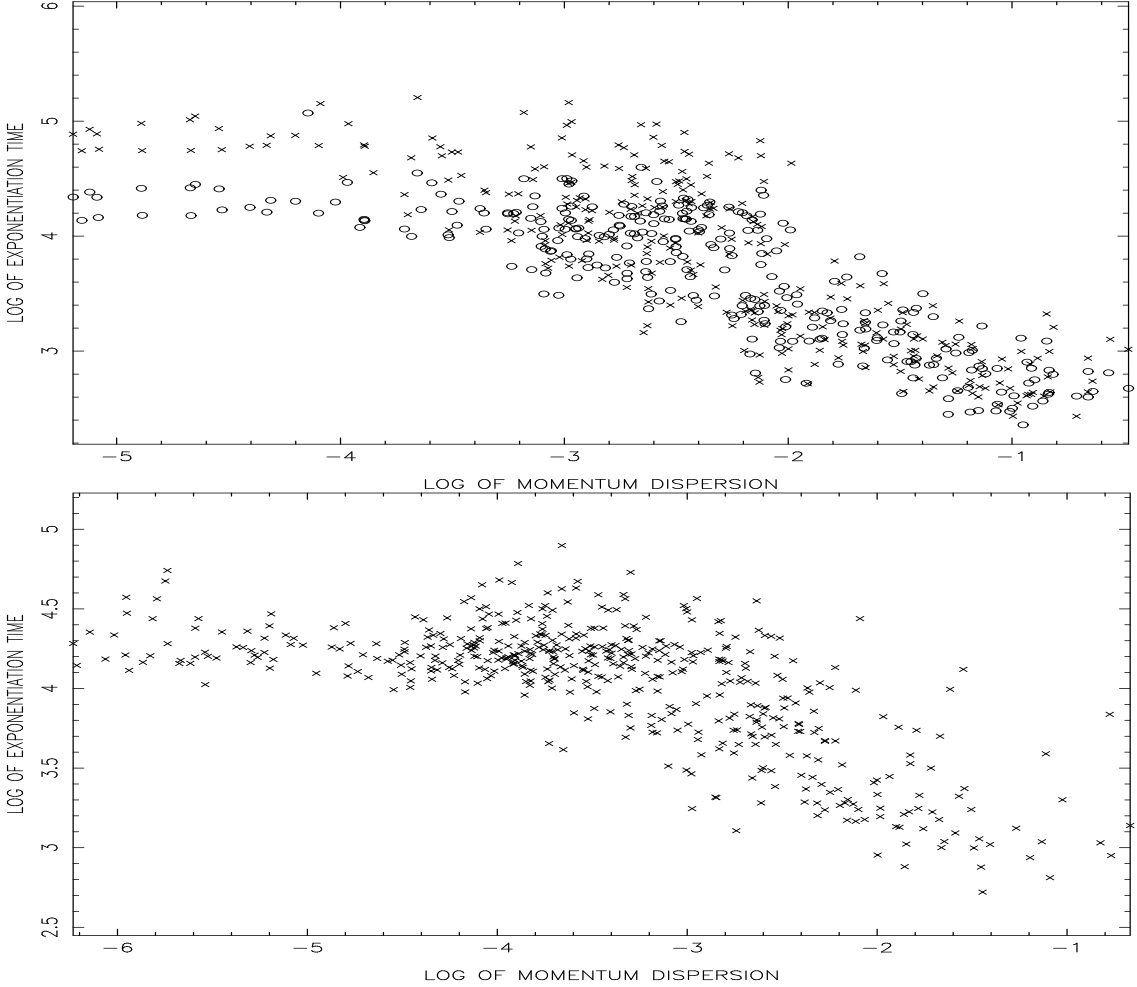

\epsfig{file=dispplot1000r2diffexp.ps,width=15.cm,height=6.5cm,angle=-90}
\epsfig{file=displot50000r2.ps,width=15.cm,height=6.5cm,angle=-90}
\caption{Relative dispersion in the angular momenta along trajectories of systems 
of $10000$ (upper panel) and $50000$ particles distributed such that 
$\rho \propto 1/r{2}$. In the upper panels the little circles correspond to 
quantities averaged over $420$ time units while the crosses correspond to
ones averaged over $840$ time units. 
 }
\label{displot}
\end{figure*}

In a smooth spherical system, the angular momenta of 
trajectories are exactly conserved.
Along with energy conservation, these guarantee the integrability of  steady
state systems, in turn ensuring that all Liapunov exponents tend to zero. The 
existence of non-zero exponents thus implies that angular momenta are not 
conserved. Moreover, if these exponents contain meaningful quantitative 
information about the 
degree as to which  dynamical properties differ from 
the smooth integrable  case, then                      
 the  conservation of the angular momenta should correlate with
the exponentiation times of our trajectories. 
In order to verify this, the angular momenta were sampled 
at intervals of one time unit, 
and the standard deviation calculated and divided by the mean. 
The results are shown
in Fig~\ref{displot}, where we have plotted the 
exponentiation times as a function of the relative dispersion in the 
total angular momenta for $N=10000$ and $N=50000$. 
 In order to examine the effect of sampling  
and to verify convergence, for the case of $N=10000$,
we have calculated the dispersions over  
$420$ (circles) and $840$ (crosses) time units (for $N=50000$ all results
are shown for trajectories integrated up to $420$ units). 
The dispersions
were found to be nearly identical in both cases, but since the Liapunov
exponents take longer time to converge to smaller values, 
there is a corresponding  increase in the e-folding times when the 
integration time was  doubled. 
Evidently, for $N=10000$ and $N=50000$, where the motion is not dominated
by discreteness noise, a clear correlation exists between the dispersion
and the e-folding times.

\begin{figure*}
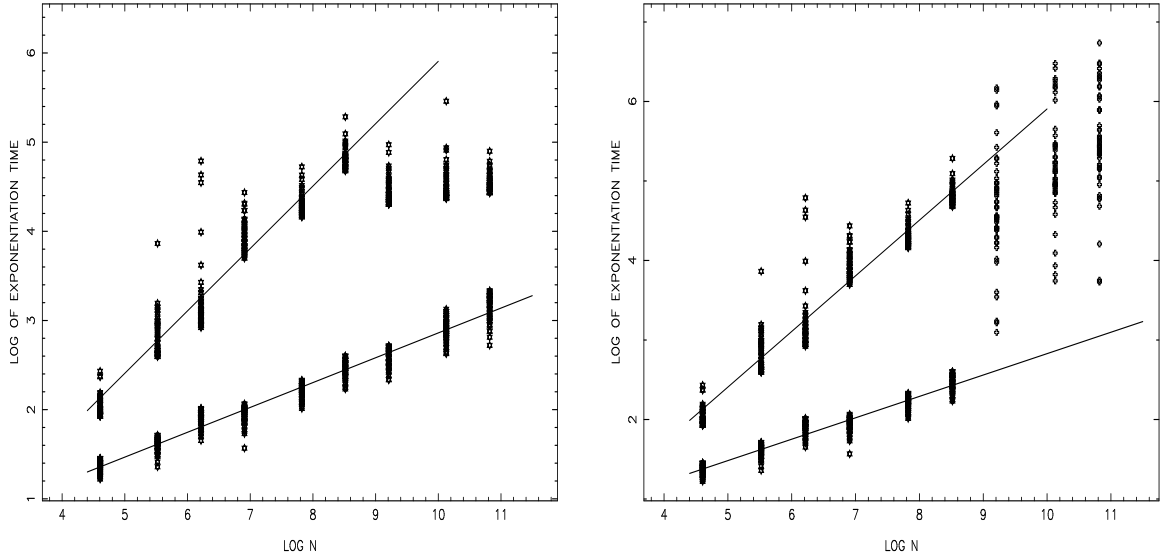

\epsfig{file= twolinesr2.ps,width=7.3cm,height=7.3cm,angle=-90}
\hspace{0.5cm}
\epsfig{file= longr2plot.ps,width=7.3cm,height=7.3cm,angle=-90}
\caption{The variation of the exponentiation times with
particle numbers for trajectories of systems of fixed particles 
distributed such as $\rho \propto 1/r^{2}$. Lines with slope 
$0.28$ are fits to the exponentiation times of trajectories 
trajectories with lowest initial angular momenta. Lines with slope
$0.72$ are best fits to trajectories with highest initial angular
momenta. For each system, out of the of $500$ trajectories integrated,
those $50$  with the highest Liapunov exponents and another fifty with 
the lowest ones are shown.  In the panel on the left all trajectories are 
integrated for $420$ time units. On the right, trajectories with the largest 
exponentiation times  are integrated for $4200$ in the cases of 
$N=10000, 25000, 50000$.
}
\label{twolinesr2}
\end{figure*}

The two sets of trajectories with markedly different Liapunov numbers 
turn out to also have different variation of these numbers as a function
of $N$. This is shown in Fig~\ref{twolinesr2}. 
The two lines drawn are least square 
fits with slopes  $0.28$ and $0.72$ ---  roughly consistent with  those 
of $N$-body systems with relatively small $N$ and trajectories in 
homogeneous distributions of non-interacting  softened particles
respectively. One
can also see that the upper line fits the data only for $N < 10000$, for
larger $N$ saturation appears to set in. Since this could be due to the 
short standard integration time of $420$ units, we integrated the 
low Liapunov exponent trajectories in systems of
$N=10000, 25000, 50000$ for up to $4200$ time units.
The results are shown in the right hand panel of 
 Fig~\ref{twolinesr2}. Although for some trajectories the 
exponentiation time-scales continue to rise 
 so as to be consistent with the growth observed for smaller values of 
$N$, some exponentiation times actually {\em decrease} --- their values becoming 
consistent with those of the  population with the smallest exponentiation times 
rather than with 
their original group! This is because, if one integrates long enough, 
trajectories can diffuse through phase space between the two 
distributions --- fluctuations can eventually lead to changes in angular 
momenta even for the more regular orbits. 
When this decreases, relatively regular trajectories can diffuse 
into the highly chaotic region. 
On the other hand, it was found that for some of the trajectories, the time dependent
Liapunov exponents
{\em continue to  decrease as the integration proceeds} (up to $4200$ units) these
 {\em are therefore candidates for being regular
trajectories}. Whether longer integrations would confirm this or whether they too 
 would eventually
escape from the regular regions (through Arnold diffusion or some other mechanism) is an 
open question. 

Clearly the phase space structure of 
these systems is highly complex and it may also well be worth further 
investigation. It is clear however that for short enough timescales, 
of the order of hundreds of crossing time, a certain 
convergence towards the regular behaviour described by the continuum limit is
apparent. This can be inferred, for example, from Fig.~\ref{momplot}, where
one can see that for $N=50000$ relatively more trajectories have exponentiation 
times approximating that of the most
regular orbits than for $N=10000$.   Moreover, this exponential time is 
 constant over a large range of initial momenta and is essentially fixed by 
the integration time --- during that time, the behaviour of the
time dependent Liapunov exponents of these trajectories indicates that
they have properties of regular trajectories (i.e., the exponents monotonously
decrease with time).

\section{Summary and concluding remarks}
\label{conremks}

In this paper the stability of the trajectories of softened gravitational
systems was examined.  The exponential instability characterising the 
divergence of nearby trajectories  in the 
case of self-gravitating systems of point particles is also
present in the corresponding softened systems.
However, whereas in the case of unsoftened $N$-body systems,
the associated e-folding times  
do not depend on the number of particles, in the case 
of softened systems, the timescales
do vary with $N$. For $N$ up to a few hundred this variation is a power law
$N^{a}$ with $a \approx 0.4$, roughly consistent with $a = 1/3$
predicted by GHH on the basis of analysing random encounters
of softened  particles in an infinite distribution. For larger $N$ (up
to a few thousand) however, the exponentiation times increase much 
more slowly. This effect appears to be connected to large scale time
dependent  variations in the potential that are swamped at small 
$N$ by discreteness noise. If this is indeed the case, then one might
expect the situation to be different if self gravity is not taken into
account .

To see if the slow rate of increase in the exponential times of $N$-body
systems did indeed arise from collective effects, 
we have integrated trajectories in centrally concentrated and homogeneous 
distributions of non-interacting particles, which were either moving independently
or were kept fixed. In the homogeneous case, not only was there no clear sign of saturation
in the $N$-dependence of the e-folding times, but the rate of increase was characterised 
by a power law with another, much larger, value of the exponent
($a \approx 0.7$) for $N$ up to $100000$.
In addition, the  
dispersion in the values of the Liapunov exponents integrated 
from different initial conditions {\em increases}
with $N$. This is indicative of a  complex phase space structure,
perhaps a mixed phase space where regular and chaotic regions coexist.
Indeed, for $N= 50000$ and $N=100000$, what appeared to be {\em completely 
regular} trajectories --- characterised by  time dependent
Liapunov exponents monotonously decreasing for thousands of dynamical
times --- were found. This phenomenon is, of course, completely absent 
in the case of point particles systems (Valluri \& Merritt 1999).

     For $N \ge 10000$,
     trajectories of systems of fixed particles that, instead of being
homogeneously distributed, had radial density profiles varying as 
$\sim 1/r^{2}$ could be classified into two groups. High angular
momentum trajectories had, in general, larger exponentiation times
than low angular momentum ones. This situation is analogous to the case
of smooth centrally concentrated 
non-spherical potentials, where eccentric trajectories can be chaotic.
It therefore appears that, in the present case, 
 discreteness noise replaces the effect of  global non-spherical
perturbations as the trigger of chaotic behaviour.

As apposed to the case of point particle systems, the 
divergence timescales of the trajectories correlate rather well with
the conservation of the action variables 
(as evidenced by the conservation of  the angular momenta) along them.
 That is, as may be expected if the exponents
tell us something about the phase space diffusion, 
more chaotic trajectories (i.e., those with larger Liapunov exponents)
conserve their action variables less efficiently
than the more regular orbits (which should approximate ones in the smoothed
out potential, where these quantities are exactly conserved).

In centrally concentrated  systems, the most chaotic trajectories and the least 
chaotic ones also have different $N$-variations of the exponential timescales,
with the former variation being a power law with  $a=0.28$ and the latter
with $a=0.72$. These values correspond to those of  
 $N$-body systems of a few hundred particles 
and of trajectories of homogeneous non-interacting distributions respectively. 
This is result is also analogous to what is found in smooth non-spherical: 
in the continuum limit, homogeneous
systems have harmonic potentials and  contain only regular orbits.
The low Liapunov exponent trajectories in the centrally 
concentrated systems also correspond to orbits that remain regular
when non-spherical perturbations are introduced.
$N$-body
systems, on the other hand, are inhomogeneous. 
The largest Liapunov exponent in an $N$-body system might then be expected 
to  correspond to that of the most
chaotic trajectories in a corresponding fixed particle system --- provided
there are no collective effects influencing the results.
This expected correspondence, and its existence in small-$N$ systems,
may be taken as further evidence that the saturation in the 
$N$-body case might be due to such effects: under the action of collective
phenomena the self consistent field is time varying, and therefore 
the saturation does not contradict the possibility that such systems
obey the CBE --- even in spherical but 
time dependent smooth potentials trajectories can be chaotic.
 Weinberg (1998) has shown that  long lived modes can be 
triggered by discreteness noise even for particle numbers far larger
than those of the  $N$-body systems described here, and can  
lead to evolutionary effects on a timescale shorter than that
of the unamplified discreteness noise 
(see also Kandrup \& Severne 1986).

   From the above discussion it is clear that the local stability of trajectories 
of softened gravitational systems can tell us much more about their dynamical
properties than in the case of systems consisting 
of point particles. Once
the effect of the singularity in the potential
is removed, many more interesting features become apparent, and obvious
correlations with some important dynamical properties are then revealed.
It thus appears that the investigation of the local stability of motion 
in this case should provide clues concerning  the mechanisms driving chaotic
behaviour in gravitational systems and their interaction.
They include high frequency  discreteness noise, larger scale collective 
phenomena and the effect of the shape of the global smoothed out potential.
These are precisely the same phenomena that drive macroscopic evolution in 
gravitational 
systems. Since, in the absence of chaotic behaviour, orbital action variables are 
conserved, once ``phase mixing'' in the corresponding angle
variables has taken place, no further evolution 
is possible. Therefore, as has been argued in El-Zant(1997), the macroscopic 
evolution of gravitational systems is necessarily driven by chaotic motion resulting
from the aforementioned mechanisms and their interactions. 
The study of the local  stability of trajectories of gravitational systems is 
thus not just interesting on intrinsic merit, but helps isolate and 
quantify evolutionary mechanisms.

Since  actual astronomical systems 
interact through the singular Newtonian potential and not a softened 
version of it, one may wonder as to the applicability of the results described
here to the orbital stability of real systems. 
Nevertheless, most studies of large dissipationless stellar systems  
assume (either explicitly or implicitly) that these systems are described 
by the CBE --- an assumption that, as pointed out in the introduction, can only,
strictly speaking, be justified in the case of softened systems.
The study of softened systems therefore is important for testing the validity
of this widely used assumption.
In this paper, as far as I am aware, first evidence is given 
of the  convergence towards the regular dynamics described by the CBE for
steady state systems with separable potentials. The fact that this convergence
is much faster in systems where self gravity has been artificially suppressed, 
and a statistical steady state forced, may be of crucial physical relevance.
Nevertheless, if one assumes, as argued in the introduction, that the mechanism
leading to the short $N$-invariant e-folding time in point particle systems
is physically unimportant, then the results presented here may be seen as an
important step in resolving the long standing apparent paradox concerning the 
exponential divergence of trajectories of gravitational systems.

In El-Zant (1997) as well as in a
couple of other papers (El-Zant 1998b; El-Zant \& Gurzadyan 1998) another,  
less direct 
(but perhaps more powerful), {\em geometric} method chracterising  
the local divergence 
of geodesics on the (Lagrangian) configuration space of dynamical systems,
was applied to dynamic and static gravitating ones.
 There too, attempts to find relations between macroscopic
evolution and microscopic instability were made, and correlations between some 
characteristic 
physical quantities (e.g., rotation, clustering and central concentration) were also found.
In the more abstract geometric setting, however, it is more difficult
to isolate the physical mechanisms at work than when applying the much more
direct method used in this paper.
Moreover, in  the case of point particles, where in the geometric case the  divergence is mainly due
to the negative curvature of the configuration manifold, the relation between the methods
is far from clear. Not only does the negative curvature
of the configuration space of point particle systems measure time independent 
deviations between
orbits, and not temporal trajectories as in the method used here, but also the terms giving 
rise to the exponential instability (on a timescale similar to that found using Liapunov exponents)
involve sums of the {\em first} derivatives of the potential, and not {\em second}
derivatives as the case when integrating the variational equations. In addition, highly technical questions on 
whether the local exponential divergence of geodesics on its own justifies the usual 
{\em global} conclusions required to infer macroscopic evolution  
(outside the idealised domain encountered in mathematical studies 
 where this was shown to be the case: see, e.g., Szczesny \& Dobrowolski 1999)
are involved. This is especially relevant since point mass systems have singular potentials,
which makes the metric also singular and the 
curvature undefined  at some points of the configuration manifold,
which then becomes ``incomplete'' (e.g., Abraham \& Marsden 1978). 
The global structure of geodesics can be affected by these singular 
``boundaries''. When, by softening the potential, the singularity is removed, 
the curvature is no longer negative. Chaotic behaviour can then result only 
from fluctuations in the curvature along the motion, the amplitude of which will decrease with
particle numbers. These fluctuations can also  be related to the Liapunov exponents
(see Casetti, Pettini \& Cohen 2000; Latora, Rapisarda \& Ruffo 1999 and the references 
therein). It is probably then in this regime  that correspondence
between the two methods can be most easilly examined.

\section*{Acknowledgments}

I would like to thank David Merritt, Adi Nusser
and  Martin Weinberg for helpful discussions and Isaac Shlosman
for commenting on the manuscript.
Much of this work was done  while the author was supported by a 
Lady Davis fellowship at the  Israel Institute of Technology.
This work was also supported by  NASA grants HST GO-08123.01-97A, 
NAG 5-10823 and WKU-522762-98-6.

\bsp

\label{lastpage}


\begin{thebibliography}{99}
\bibitem{}
Abraham R., Marsden J.E., 1978, Foundations of mechanics. Benjamin, Reading, Massachusetts
\bibitem{}
Antonov V.A., 1962, Vestn. Leningrad Gos. Univ. 7,135 (English translation In:
Goodman J., Hut P. (eds.) Dynamics of globular clusters, IAU symp. 113. 
Reidel, Dordrecht, 1985
\bibitem{}
Benettin G., Galgani L., Strelcyn J.M., 1976, Phys. Rev. A14, 2338
\bibitem{}
Braun W., Hepp K., 1977, Commun. Math. Phys. 56, 101
\bibitem{}
Caranicolas N.D., Innanen K.A., 1991, AJ 102, 1343
\bibitem{}
Casetti L., Pettini M., Cohen E.G.D., 2000, Phys. Rep. 3, 237  
\bibitem{}
El-Zant, A.A., 1997, A \& A 326, 113
\bibitem{}
El-Zant A.A., 1998a, Phys. Rev. E58, 4152
\bibitem{}
El-Zant A.A., 1998b, A \& A 331, 782
\bibitem{}
El-Zant A.A., Gurzadyan V.G., 1998, Physica D122, 241
\bibitem{}
El-Zant A.A., Hassler B., 1998, New A. 3, 393
\bibitem{} 
Gerhard O.E., Binney J., 1985, MNRAS 216,467
\bibitem{}
Goodman   J., Heggie D.C., Hut P., 1993, ApJ 415, 715 (GHH)
\bibitem{}
Kandrup H.E., Severne G., 1986, Ap \& SS 126, 177 
\bibitem{}
Kandrup H.E., Mahon M.E., Smith H., 1994, ApJ 428, 458
\bibitem{}
Latora V., Rapisarda A., Ruffo S., 1999, Physica A273, 97
\bibitem{}
Lichtenberg A.J., Lieberman M.A., 1983, Regular and stochastic motion.
Springer, New York 
\bibitem{}
Lynden-Bell D. and Wood R., 1968, MNRAS 138, 495
\bibitem{}
Merritt D., Fridman T., 1996, 440,136
\bibitem{}
Miller R.H., 1964, ApJ 140, 250 
\bibitem{}
Padmanabhan T., 1990, Phys. Rep. 188, 285
\bibitem{}
Schwarzschild M., 1993, ApJ 409,563
\bibitem{}
Szczesny J.  Dobrowolski T., 1999, (Chao-dyn/9901007)
\bibitem{}
Spohn H., 1980, Rev. Mod. Phys. 52, 569
\bibitem{}
Valluri M., Merritt D., 1999, Orbital instability and relaxation in stellar 
systems. In: 
 Ruffini R.,  Gurzadyan V.G. (eds.)
The Chaotic Universe.
World Scientific (astro-ph/9909403)
\bibitem{}
Weinberg M. D., 1998, MNRAS 297, 101
\bibitem{}
Wiggins S., 1991, Chaotic transport in dynamical systems. 
Springer, New York
\bibitem{}
Wolf A., Swift J.B., Swinney H.L., Vastano J.A., 1985, Physica 17D,288
\end{thebibliography}
\end{document}